# EAST Real-Time VOD System Based on MDSplus

J.Y. Xia, B.J. Xiao, Fei Yang and Dan Li

*Abstract*—As with EAST (Experimental Advanced Superconducting Tokamak） experimental data analyzed by more and more collaborators, the experimental videos which directly reflect the real status of vacuum attract more and more researchers' attention. The real-time VOD（Video On Demand）system based on MDSplus allows users reading the video frames in real time as same as the signal data which is also stored in the MDSplus database. User can display the plasma discharge videos and analyze videos frame by frame through jScope or our VOD web station. The system mainly includes the frames storing and frames displaying. The frames storing application accepts shot information by using socket TCP communication firstly, then reads video frames through disk mapping, finally stores them into MDSplus. The displaying process is implemented through B/S(Browser/Server) framework, it uses PHP and JavaScript to realize VOD function and read frames information from MDSplus. The system offers a unit way to access and backup experimental data and video during the EAST experiment, which is of great benefit to EAST experimenter than the formal VOD system in VOD function and real-time performance.

## I. INTRODUCTION

EAST experimental data is shared and analyzed by more and more collaborators in recent years, especially the experimental videos which directly reflect the real status of inner vacuum. Therefore, we built an EAST VOD system in the early time. However, as with more and more high-speed cameras are installed in the EAST machine, the original VOD system need to be improved in video storing and displaying to make the system more user-friendly and time-saving.

MDSplus [1] is a set of software tools for data acquisition and storage which is already used in EAST experiment. MDSplus allows all data from an experiment or simulation code to be stored into a single, self-descriptive, hierarchical structure.

Therefore, the new EAST VOD system based on MDSplus is built, storing the video frames into MDSplus, allowing users read the video frames in real time as same as the signal data which is also stored in the MDSplus database, it is of great benefit for EAST researchers.

In this paper, the advantages of the improved EAST VOD system are introduced and the technical details are presented. In section II, we discuss the new system architecture. Section III shows the acquisition and storage process before the video displaying. Section IV introduces the detailed realization of the VOD function. In section V, the improvements of new system are described comparing with the formal EAST VOD system. Finally, we give a conclusion and discuss the future work in the last section.

## II. SYSTEM ARCHITECTURE

The architecture of the real-time VOD system based on MDSplus is shown as Fig.1，which is composed of center control server, acquisition server, storage server and the VOD server.

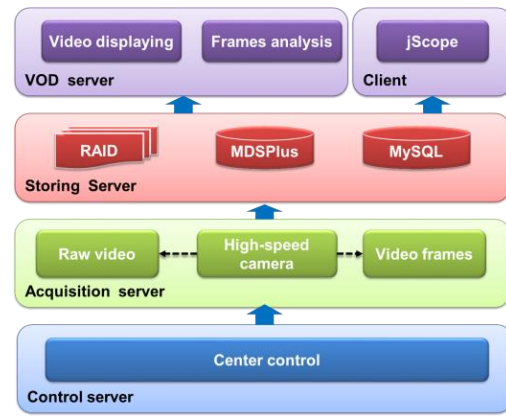

Fig.1. The architecture of the new EAST VOD system.

*Center control system*

The central control system(CCS) is the core of EAST experiment's management system to coordinate all of subsystems according to pre-set value and logic[2]. The center control system will trigger the high-speed cameras to start work when the EAST experiment is beginning.

*Acquisition server*

Then the cameras start to record the raw experimental video, split the video to image frames and save timestamps into a file. Meanwhile it sends the "shot number" to the storage server throw TCP socket communication.

This work is supported by National Key R&D Program of China (Grant No: 2017YFE0300500, 2017YFE0300504).

J.Y. Xia is with the institute of Plasma Physics, Chinese Academy of Sciences, Hefei 230031, China (e-mail: jyxia@ipp.ac.cn).

B.J. Xiao is with the institute of Plasma Physics, Chinese Academy of Sciences, Hefei 230031, China (e-mail: bxiao@ipp.ac.cn).

Fei Yang is with Department of Medical Informatics Engineering, Anhui Medical University, Hefei 230031, China (e-mail: fyang@ipp.ac.cn).

Dan Li is with the institute of Plasma Physics, Chinese Academy of Sciences, Hefei 230031, China (e-mail: lidan@ipp.ac.cn).

*Storing server*

The videos are stored in the server with RAID（Redundant Array of Inexpensive Disks）, the frames and timestamps are saved into MDSplus, and the shot information for webpage is stored in MySQL database.

*VOD server*

Finally, the VOD server reads frames from MDSplus and displays the video and frames on the web browser

### III. Data acquisition and Storage

There are four kinds of cameras used in the EAST experiment, including one infrared camera and three visible CCD. The infrared diagnosis is installed in windows K and used to measure the temperature evolution and distribution on divert or plates, with 47°*59° view field and 320*240 resolution [3]. The other three visible CCD monitor the window D, window G and window K through a wild-angle endoscope for visible light observation [4]. The cameras start to acquire the video after receiving the central control's pulse signal. When one shot finish, the video acquisition stopp and the raw video will be saved in the storage server with RAID. The size of raw video is very large and need specific video player to play, so that it is impossible to display the raw video on the internet. Therefore we split the raw video to frames with a certain sampling rate and save timestamp into *.txt file. The frames and time file are save in the shared directory, and the acquisition server sends the shot number to the storage server through the TCP socket communication.

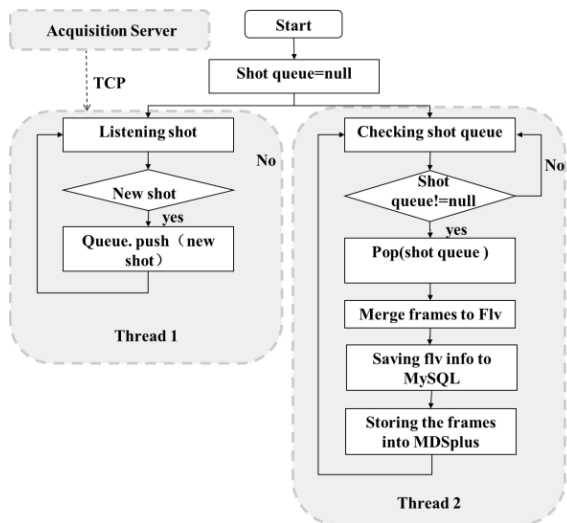

Fig.2. The flow chart of data storage.

The storage server receives the shot number and starts to storing the frames and videos through two thread as shown in Fig.2.

*Thread1*

Monitoring and accepting the shot massage from acquisition server through TCP communication, then saving the shot number into a queue;

*Thread2*

Checking the shot queue, if the shot queue is not null, the front shot number will be read and start read the frames and time file by disk mapping. Then, the frames will be used to store synthetic video through FFmpeg [5], which is a complete, cross-platform solution to synthetic the frames to a *.flv video. Then we store videos and shot information into database.

On the other hand, MDSplus segments are useful to store sequences of images (frames) into MDSplus.

### IV. Displaying Process

The Video on demand system based on MDSplus is a website including video displaying webpage as shown in Fig.3 and frame analysis webpage as shown in Fig.4.

The video displaying webpage is composed of menu bar，search bar, shot list and video displaying panel. User can select different kind of videos through the menu bar. On one hand user can directly display or download the video of one shot in the shot list, on the other hand user can display the video of a certain shot by inputting the shot number in search bar.

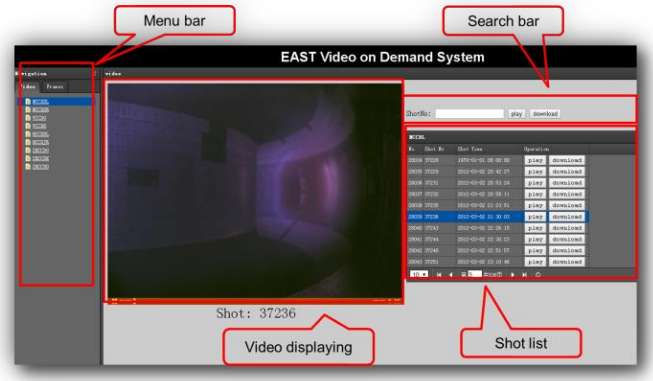

Fig.3. Video displaying webpage

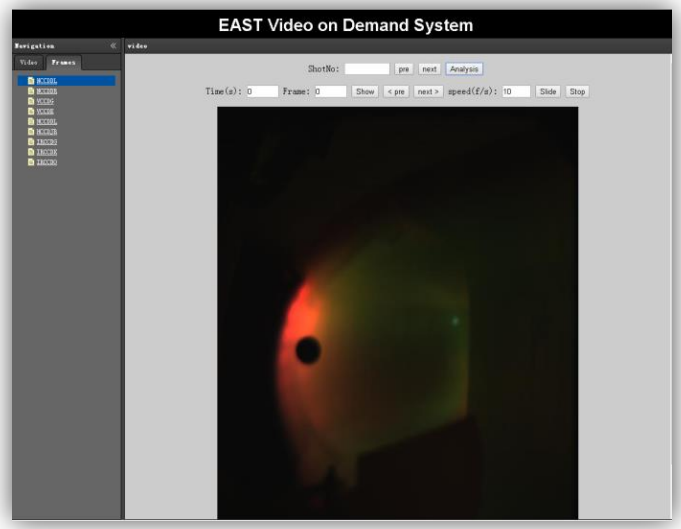

Fig.4. Frame analysis webpage

The frame analysis webpage is realized by reading frames and timestamps from MDSplus database, it allows user viewing the video frame by frame according to the sample ratio users setting. User can analysis the frames of a certain shot by setting shot number, and setting the speed of frame displaying, meanwhile user can get the specific time of frame through the "pre", "next" and "show" button.

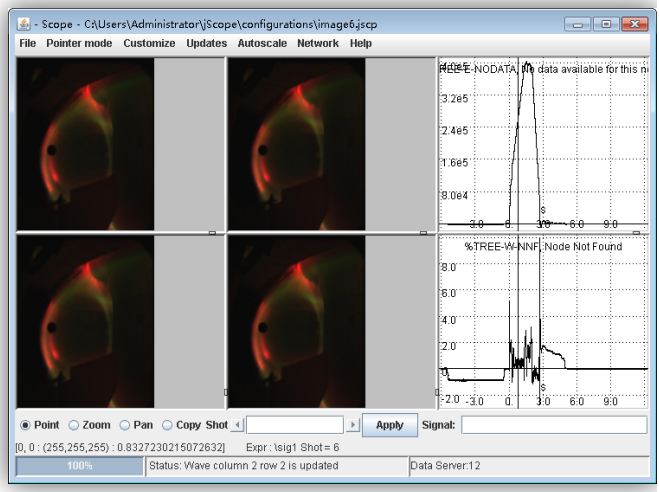

Fig.5. Video displaying in jScope

jScope is a MDSplus tool for quick waveform displaying during machine operation [6]. The new version of jScope allows displaying the content of the sequence of frames. Therefore we can read several EAST experimental frames along with various experimental signal data from MDSplus as shown in Fig.5. User can play the frames animation and data waveform animation at the same time through "Pointer mode".

## V. SYSTEM COMPARISION

The old system saves the image frames in RAID and save the timestamp in a *.xml file in windows system. The new system saves the image frames and timestamp in MDSplus database in Linux system. According to the real-time comparison as shown in TABLE I, the new system is nearly three times faster than the old system, especially the long pulse shot in EAST experiment, such as "73999" .

TABLE I
REAL-TIME COMPARISON

| Shot No | Length (s) | Size (MB) | Frames | Old time (s) | New time (s) |
|---|---|---|---|---|---|
| 77212 | 0.78 | 1.02 | 97 | 7.9735 | 1.0145 |
| 77213 | 9.79 | 30.9 | 1198 | 30.002 | 10.518 |
| 77214 | 5.31 | 15.3 | 650 | 16.787 | 4.896 |
| 77215 | 9.37 | 31.7 | 1146 | 26.226 | 9.368 |
| 77216 | 9.48 | 31.8 | 1160 | 29.376 | 8.778 |
| 73999 | 104.78 | 169.8 | 12810 | 271.607 | 91.875 |

## VI. CONCLUSION

Eventually, the EAST real-time VOD system based on Mdsplus is an user-friendly system. On one hand, it offers a more highly-interactive interface for scientists to view the plasma discharge videos and analysis the video frame by frame. On the other hand, it allows users to store and read the frames through the same way as experimental data through MDSplus.

Compared with the former system, the new system shows some optimization and improvement in expansibility and real-time.

The next step will be to enhance the VOD function and improve real timing, for example, adding the function of various videos synchronous display and reducing the time of video synthesis..